\documentclass{Interspeech}

\usepackage{amsmath,amssymb,amsfonts}
\usepackage{algorithmic}
\usepackage{graphicx}
\usepackage{textcomp}
\usepackage{xcolor}
\def\BibTeX{{\rm B\kern-.05em{\sc i\kern-.025em b}\kern-.08em
    T\kern-.1667em\lower.7ex\hbox{E}\kern-.125emX}}
\usepackage{soul,framed} %,caption
\colorlet{shadecolor}{yellow}
\usepackage{array}
\usepackage{url}
\usepackage{cite}
\usepackage{booktabs}
\usepackage{multirow}
\usepackage{multicol}
\usepackage{float}
\usepackage{bbold}
\usepackage{enumitem}
\usepackage{hyperref}
\usepackage{etoolbox,siunitx}
\robustify\bfseries
\usepackage{balance}
\usepackage{pifont}
\newcommand{\cmark}{\ding{51}}%
\newcommand{\xmark}{\ding{55}}%

\usepackage{scalerel}

\interspeechcameraready 
\title{Online Audio-Visual Autoregressive Speaker Extraction}

\author[affiliation={1}]{Zexu}{Pan}
\author[affiliation={2}]{Wupeng}{Wang}
\author[affiliation={1}]{Shengkui}{Zhao}
\author[affiliation={1}]{Chong}{Zhang}
\author[affiliation={1}]{Kun}{Zhou}
\author[affiliation={1}]{Yukun}{Ma}
\author[affiliation={1}]{Bin}{Ma}

\affiliation{Tongyi Lab}{Alibaba Group}{Singapore}
\affiliation{}{National University of Singapore}{Singapore}
\email{zexu.pan@alibaba-inc.com}

\keywords{Cocktail party problem, real-time, speaker extraction, visual frontend, autoregressive}

\usepackage{comment}

\begin{document}

\maketitle

\begin{abstract}
This paper proposes a novel online audio-visual speaker extraction model. In the streaming regime, most studies optimize the audio network only, leaving the visual frontend less explored. We first propose a lightweight visual frontend based on depth-wise separable convolution. Then, we propose a lightweight autoregressive acoustic encoder to serve as the second cue, to actively explore the information in the separated speech signal from past steps. Scenario-wise, for the first time, we study how the algorithm performs when there is a change in focus of attention, i.e., the target speaker. Experimental results on LRS3 datasets show that our visual frontend performs comparably to the previous state-of-the-art on both SkiM and ConvTasNet audio backbones with only $0.1$ million network parameters and $2.1$ MACs per second of processing. The autoregressive acoustic encoder provides an additional $0.9$ dB gain in terms of SI-SNRi, and its momentum is robust against the change in attention.
\end{abstract}

\section{Introduction}
\label{sec:introduction}

Real-world speech signals are often mixed with interfering speech and noise signals. Human brains excel at focusing on an interested speech signal, i.e., target speech, while filtering out the rest, known as auditory attention~\cite{cherry1953some}. Equipping machines with such auditory selective attention is crucial for speech applications such as automatic speech recognition~\cite{Wang_aaai_2024}.

Speech separation algorithms separate a multi-talker speech signal into individual streams by speakers~\cite{hershey2016deep,luo2019conv,luo2020dual,wang2023tf,zhao2023mossformer2}. However, the separated speech signals are not associated with specific speaker identities. In contrast, speaker extraction research exhibits the attention to disentangle only a target speaker's speech signal from the multi-talker speech signal, with the attention driven by an auxiliary conditioning signal. The auxiliary signal could be a pre-recorded speech signal for the network to attend to a speech that has a similar voice signature~\cite{wang2019voicefilter,vzmolikova2019speakerbeam,Chenglin2020spex}, or a face recording for the network to attend to the synchronized speech~\cite{ephrat2018looking,pan2023avse,wu2019time,pan2021reentry,chung2020facefilter}, or even the brain signal for the network to attend to the speech that the brain is trying to listen~\cite{biss2020,pan2023neuroheed,pan2024neuroheed+}.

This research focuses on audio-visual speaker extraction (AVSE) in online scenarios, addressing the need for streaming on-device AVSE algorithms in applications like video conferencing and human-robot interactions, such as in-car or service robot communication. Due to device constraints, on-device models usually favor fewer network parameters and multiply-accumulate (MAC) operations, to accelerate the processing, which also enhances the real-time factor. While most existing research emphasizes optimizing the audio component of the network~\cite{wang2020voicefilter,papez2023,li2022skim,2024resepformer,chen2024rt}, limited studies have focused on improving the visual counterpart. The visual encoder is still predominantly based on ResNet18~\cite{usev21,chen2024rt}, with recent explorations into ShuffleNet-based networks~\cite{Zhu2023rt}. In this work, inspired by the effectiveness of the BlazeFace face detection network~\cite{blaze2019} and depth-wise separable convolution~\cite{xception2017}, we propose a lightweight visual encoder that reduces feature dimension size while incorporating deeper depth-wise separable convolutional layers to encode lip motion more efficiently.

We also aim to enhance the audio components of online AVSE. During real-time inference, while the network processes newly acquired frames of a mixture signal, it also has access to the separated speech signals from previous frames. This availability makes autoregressive networks well-suited to leveraging this past information. 
In the literature, the streaming NeuroHeed model, which conditions on brain signals, encodes the separated signals into a single speaker embedding vector using a speaker encoder and repeatedly fuses it with every speech frame in its sliding window-based decoding process~\cite{pan2023neuroheed}. The PARIS speech separation research also explored incorporating past separated speech frames into the mixture signal inputs to utilize historical information in frame-level decoding~\cite{pan2024autoregressive}. In this work, we adopt frame-level decoding like PARIS, which is better suited for online decoding with improved real-time performance, differently, we propose encoding the past extracted signals with a lightweight acoustic encoder, to generate distinct acoustic embeddings at the frame level for the speaker extractor, which serves as a secondary conditioning signal alongside the visual cue.

Another common scenario in real-world deployments of streaming AVSE models is the changing focus between active speakers, such as in multi-party meetings and conversations. The network must adjust its attention dynamically to extract the correct speech signal. Previous works assume the identity of the target speaker is consistent. In this work, we explore the changing target scenario for the first time, demonstrating that visual-conditioned speaker extraction models exhibit robust performance when faced with such changes in attention, which is crucial for real-world applications. Furthermore, our proposed acoustic encoder, which encodes past information into frame-level conditioning embeddings, shows robustness in adapting to changes in attention. This contrasts with the NeuroHeed model, where aggregating all information into a single speaker embedding may create momentum that hinders the network's ability to adjust focus quickly.

The contributions of this work for online AVSE studies are threefold: 
1) We propose a lightweight visual encoder for AVSE, which, with only $0.1$ million parameters and $2.1$ MACs per second of processing, performs comparably to the previous state-of-the-art visual encoders on both SkiM and ConvTasNet audio backbones.
2) We introduce an autoregressive acoustic encoder for streaming AVSE, which provides an additional $0.9$ dB gain in SI-SNRi on the Lip Reading Series 3 (LRS3) dataset~\cite{afouras2018lrs3}.
3) We conduct the first study on the switching attention scenario, demonstrating that our proposed model is robust in such contexts.

\section{Proposed network}
Let $x(\tau)$ be a multi-talker mixture speech signal\footnote{Variables with $\tau$ denote speech signals in the time domain, while variables with $t$ denote frame-based embeddings.}, consisting of the target speech signal $s(\tau)$ and interference speech signal $b(\tau)$, the AVSE network $f(\cdot)$ estimates the target speech signal $\hat{s}(\tau)$ to approximate $s(\tau)$, conditioned on the visual recording of the target speaker $v(t)$:

\begin{equation}
\label{eqa:network_flow}
    \hat{s}(\tau) = f(x(\tau), v(t))
\end{equation}

In this work, we additionally study the changing target scenario, meaning that the identity of $v(t)$ and $s(\tau)$ may change in a processing clip. For simplicity, we maintain the same representation of $v(t)$ in this changing target scenario.

\subsection{Architecture}
\begin{figure}[t]
  \centering
  \includegraphics[width=0.99\linewidth]{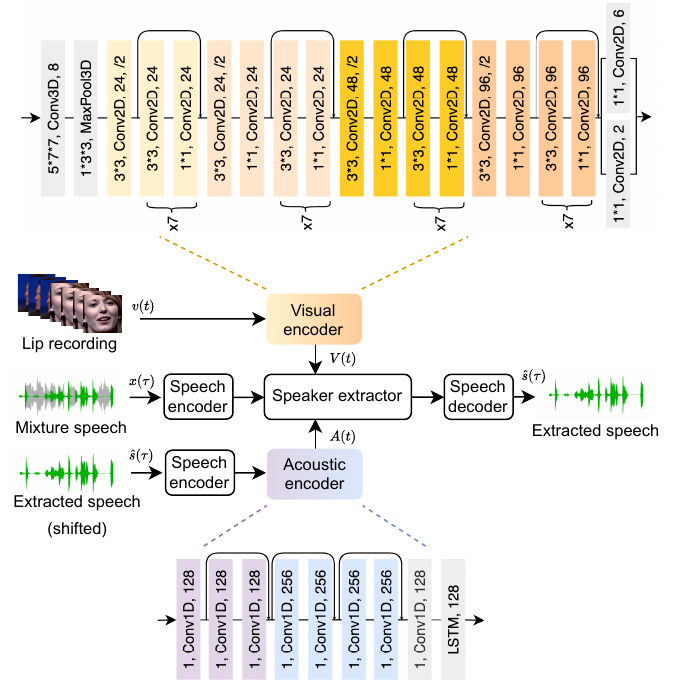}
  \vspace{-2mm}
  \caption{Our proposed online audio-visual autoregressive speaker extraction network. The middle section introduces the network components and data flow, the top section showcases our proposed lightweight visual encoder, and the bottom section depicts our proposed lightweight autoregressive acoustic encoder. Each colored layer's parameters are specified in terms of kernel size, layer type, output channel size, and stride. 
  }
  \vspace{-4mm}
  \label{fig:network}
\end{figure}

Our proposed online audio-visual autoregressive speaker extraction network is illustrated in Fig.~\ref{fig:network}. It comprises two identical non-shared-weight speech encoders, a speaker extractor, a speech decoder, a novel visual encoder, and a novel autoregressive acoustic encoder.

% \subsubsection{Speech encoder and decoder}
For speech encoder and decoder, we follow the masking-based time-domain approach~\cite{luo2019conv,li2022skim,zhao2023mossformer2}. The speech encoder includes a 1-dimensional (1D) convolution $Conv1D$ and a rectified linear activation. The speech decoder consists of a linear layer and an overlap-and-add operation. The channel, kernel, and stride sizes are set to $128$, $16$, and $8$ ,respectively.

% \subsubsection{Speaker extractor}
For the speaker extractor, we adopt the online skipping memory long short-term memory network (SkiM), which is known for its low latency and excellent performance for online speech separation~\cite{li2022skim}. We set the number of output streams to one to suit the speaker extraction task. We set the long short-term memory (LSTM) unit size to $384$, the number of layers to $3$, and the non-overlapping segment size to $50$.

% \subsubsection{Visual encoder}
Our proposed lightweight visual encoder, named BlazeNet64, is shown in the top section of Fig.~\ref{fig:network}, it encodes the lip image sequence $v(t)$ to visual cue, which is used as an attention attractor to extract the target speech. The architecture is motivated by the effectiveness of the BlazeFace face detection network~\cite{blaze2019} and depth-wise separable convolution~\cite{xception2017}. $v(t)$ is first processed by a causal 3-dimensional convolution $Conv3D$. The rest of the layers in the visual encoder are depth-wise separable 2-dimensional convolutions applied to each image independently. Compared to the widely used $ResNet18$-based visual encoder, which was originally designed for lip reading~\cite{Afouras18b}, our visual encoder is narrower but deeper, and much more light-weighted. We do not pre-train this visual encoder, so we do not use adapter layers after the visual encoder as in~\cite{wu2019time,pan2020muse}. To align with the audio embeddings, the visual embeddings $V(t)$ are temporally repeated to account for frame rate differences before concatenation with the audio embeddings at the beginning of the speaker extractor.

% \subsubsection{Acoustic encoder}
Our proposed lightweight autoregressive acoustic encoder leverages extracted clean speech samples from past frames to maintain temporal attention momentum during online processing, shown in the bottom section of Fig.~\ref{fig:network}. The extracted speech samples from the past frames will first pass through a speech encoder, and then through a series of 1D-convolutional layers before being summarized by a single LSTM layer. The output $A(t)$ is also concatenated with the audio embeddings at the start of the speaker extractor.

\subsection{Training strategy and loss function}
The network is non-autoregressive if the acoustic encoder is dropped from Fig.~\ref{fig:network}, in this case, the negative scale-invariant signal-to-noise ratio (SI-SNR)~\cite{le2019sdr} is adopted as the loss function for training:
\begin{equation}
    \label{eqa:loss_SI-SNR}
    \mathcal{L}_{\text{SI-SNR}}(s, \hat{s}) = - 20 \log_{10} \frac{|\frac{<\hat{s},s>}{|s|^2}s|}{|\hat{s} - \frac{<\hat{s},s>}{|s|^2}s\big|}.
\end{equation}

When our proposed acoustic encoder is integrated into the overall network, the network operates in an autoregressive mode. In this case, we adopt the Pseudo-AutoRegressIve Siamese Training (PARIS) strategy~\cite{pan2024autoregressive}, which involves two forward passes for each batch during training. The first pass forward the network without the acoustic encoder (with $A(t)$ being set to zero embeddings as a placeholder). The second pass performs a full forward pass with the input of the acoustic encoder being the output of the first pass which is used as the ``pseudo past extracted speech''. 

The PARIS paper suggested using the scale-sensitive signal-to-noise ratio (SNR) rather than the widely used SI-SNR as the loss function for training, to normalize network output for feedback in step-by-step streaming settings. However, networks trained with the SNR loss function typically do not perform as well as those trained with the SI-SNR loss function. Therefore, in this work, we pair SI-SNR with a scale-variant frequency domain loss~\cite{pan2022hybrid}, specifically the frequency-domain multi-resolution delta spectrum loss $\mathcal{L}_{\text{freq}}$~\cite{pan2022hybrid}, as the loss function. The loss functions for the fist pass $\mathcal{L}_{1}$ and the second pass $\mathcal{L}_{2}$ are defined as:
\begin{equation}
    \label{eqa:loss_1}
    \mathcal{L}_{1} = \mathcal{L}_{\text{SI-SNR}}(s, \hat{s}^{1}) + 0.25*\mathcal{L}_{\text{freq}}(s, \hat{s}^{1}) 
\end{equation}
% and
\begin{equation}
    \label{eqa:loss_2}
    \mathcal{L}_{2} =  \mathcal{L}_{\text{SI-SNR}}(s, \hat{s}^{2}) + 0.75*\mathcal{L}_{\text{freq}}(s, \hat{s}^{2}) 
\end{equation}
where $\hat{s}^{1}$ and $\hat{s}^{2}$ are the network outputs of the first and second pass respectively.

\section{Experimental setup}

\subsection{Dataset}
We mainly use the Lip Reading Sentences 3 (LRS3) dataset to validate our proposed method in this work~\cite{afouras2018lrs3}, which is widely used in many AVSE studies~\cite{Hsu_2023_CVPR,liu2023piave,martel2023}. The speech signal is available at $16$kHz, and video is available at $25$ frames per second.

We study the following three scenarios in this paper: 1) Two speakers speak simultaneously, with one being the target speaker. 2) Two speakers speak simultaneously, with the target speaker switching from one to the other at a random time. 3) One speaker speaks for a duration, followed by another speaker, both of whom are target speakers. An additional interference speaker overlaps with the target’s speech. 

To create the mixture utterances, we first select one utterance as the anchor. The remaining utterances are normalized to match the energy level of the anchor. Each utterance is then assigned a random SNR ranging from $10$dB to $-10$dB relative to the anchor before being combined. We simulate $40,000$, $5,000$, and $3,000$ such two-speaker overlapping mixture utterances for our train, validation, and test sets using the original LRS3 datasets. The speaker and utterances in the test set do not overlap with the train and validation sets to ensure speaker-independent analysis.

\subsection{Optimization}
We use the PyTorch framework to conduct our experiments~\footnote{The model and training scripts are available at \url{https://github.com/modelscope/ClearerVoice-Studio}}. All models trained in this work adopt the same optimization setting. We used distributed data-parallel training with four Tesla \mbox{16 GB} V100 GPUs, the effective batch size is $16$. We train the models for $150$ epochs, the Adam~\cite{kingma2015adam} optimizer is used with an initial learning rate of $0.001$. The learning rate is reduced by half if the best validation loss (BVL) does not improve for $6$ consecutive epochs, with training stopping early if the BVL does not improve for $20$ consecutive epochs. 
% Audio clips are truncated to $6$ seconds during training to fit GPU memory, while full audio clips are used for evaluation.

\section{Results}
To evaluate the quality of extracted speech, we use several metrics: the improvement in SI-SNR (SI-SNRi)~\cite{le2019sdr}, the improvement in SNR (SNRi)~\cite{vincent2006performance}, the improvement in Perceptual Evaluation of Speech Quality (PESQi)~\cite{rix2001perceptual}, and the improvement in Short-Term Objective Intelligibility (STOIi)~\cite{taal2010short}. All improvements are calculated relative to the unprocessed multi-talker speech signals. Higher values indicate better performance. We use SI-SNRi as our main metric when comparing the results, as other metrics show similar trends. For clarity, each differently trained system is assigned a unique system number (Sys).

\subsection{Comparison on visual encoders}
We first compare our visual encoder with baseline visual encoders in Table~\ref{tab:param_mac} and Table~\ref{tab:v_enc}. We consider two baselines: the ResNet18-based visual encoder and the ShuffleNetV2-based visual encoder, both are state-of-the-art approaches.

In Table~\ref{tab:param_mac}, we examine the computational efficiency. Our BlazeNet64 visual encoder stands out with only $0.1$ million network parameters, which is significantly smaller compared to the $0.9$ million and $11.2$ million parameters of the ShuffleNetV2 and ResNet18-based visual encoders, respectively. Moreover, the BlazeNet64 visual encoder requires considerably less computation, with only $2.1$ billion MACs per second of processing, in contrast to the $7.3$ billion and $12.9$ billion MACs required by the ShuffleNetV2 and ResNet18-based encoders, respectively.

In Table~\ref{tab:v_enc}, we compare the performance of our visual encoder with the baselines on the simulated LRS3 dataset. In systems 1 to 3, where AV-ConvTasNet~\cite{luo2019conv,pan2020muse} is used as the speaker extractor, the ResNet18 visual encoder achieves the best overall SI-SNRi of $9.2$ dB. Our visual encoder closely matches the ResNet18 visual encoder with an SI-SNRi of $9.1$ dB, and surpasses the ShuffleNetV2 visual encoder by $0.3$ dB. The AV-ConvTasNet speaker extractor shows suboptimal performance in the switching target scenario, with a notable drop in SI-SNRi after the switch. For example, system 3 has an SI-SNRi `after' switching of $8.5$ dB, which is $0.7$ dB lower than the SI-SNRi `before' switching.

In systems 4 to 6, where AV-SkiM~\cite{li2022skim} is used as the speaker extractor, the ResNet18 visual encoder achieves an SI-SNRi of $9.0$ dB, while our visual encoder achieves a similar SI-SNRi of $9.1$ dB. Notably, the AV-SkiM speaker extractor maintains consistent performance `before' and `after' the target switch in the switching target scenario, with similar SI-SNRi values. Across systems 1 to 6, the SI-SNRi in scenarios `without' a target switch is generally higher than in scenarios with a target switch, due to the longer audio clips providing more context for speaker extraction.

In systems 7 and 8, we evaluate our visual encoder against the ResNet18-based visual encoder under non-causal (offline) conditions. Both systems achieve an average SI-SNRi of $12.8$ dB, which is significantly higher than the performance of causal models. In systems 9 and 10, we test the impact of reducing the visual frame rate to $12.5$ and $5$ frames per second (FPS), respectively. While this reduction leads to lower computational costs, the performance drops significantly, indicating that the reduced frame rates are not worthwhile given the substantial loss in performance.

\begin{table}
    \centering
    \sisetup{
    detect-weight, % Make siunitx detect align bold cells correctly
    mode=text, % Make siuntix print tables in text mode (causes width of bold characters to be the same as non-bold)
    tight-spacing=true,
    round-mode=places,
    round-precision=2,
    table-format=1.2
    }
    \caption{Comparison of visual encoders. Parameters (Param) are reported in million (m), and multiply-accumulate (MAC) operations is in billion (G) per $1$ second input, the lower the better.}
    \vspace{-2mm}
    % \addtolength{\tabcolsep}{-3.5pt}
    \resizebox{0.8\linewidth}{!}{
    \begin{tabular}{c *{2}{S[round-precision=1,table-format=2.1]}}
       \toprule
        Visual Encoder      &{Params (M)}        &{MACs (G)}\\
        \midrule
        ResNet18            &11.18              &12.9\\
        ShuffleNetV2          &0.87               &7.3\\
        BlazeNet64 (Ours)        &0.13               &2.1\\
        \bottomrule
    \end{tabular}
    }
    % \addtolength{\tabcolsep}{3.5pt}
    \vspace*{-3mm}
    \label{tab:param_mac}
\end{table}

\begin{table*}
    \centering
    \sisetup{
    detect-weight, % Make siunitx detect align bold cells correctly
    mode=text, % Make siuntix print tables in text mode (causes width of bold characters to be the same as non-bold)
    tight-spacing=true,
    round-mode=places,
    round-precision=2,
    table-format=1.2
    }
    \caption{Comparison of our visual encoder with baselines on the LRS3 dataset. We compare its effectiveness with different speaker (spk.) extractor backbones, causal or non-causal settings, and different visual frame rates (V. FPS). We first report the SI-SNRi in dB, SNRi in dB, PESQi, and STOIi for `all' the clips, we then report the SI-SNRi in dB for utterance segments `before' the target switch, `after' the target switch, and utterances `without' target switch. The acoustic encoder is not used in these systems.}
    \vspace{-2mm}
    \addtolength{\tabcolsep}{-1.5pt}
    \resizebox{\linewidth}{!}{
    \begin{tabular}{ccccc *{2}{S[round-precision=1,table-format=2.1]} SS *{5}{S[round-precision=1,table-format=2.1]}}
       \toprule
        \multirow{2}*{Sys}      &\multirow{2}*{Spk. Extractor}   &\multirow{2}*{Causal} &\multirow{2}*{V. FPS} &\multirow{2}*{V. Encoder}  &{SI-SNRi}   &{SNRi}   &{PESQi}   &{STOIi} &{SI-SNRi} &{SI-SNRi} &{SI-SNRi}     &{\multirow{2}*{Params}}        &{\multirow{2}*{MACs}}\\
        &&&&&{(all)}   &{(all)}   &{(all)}   &{(all)} &{(before)} &{(after)} &{(without)}\\
        \midrule
        1   &\multirow{3}*{AV-ConvTasNet} &\multirow{3}*{\cmark} &\multirow{3}*{25} &ResNet18  
        &9.2233     &9.5232     &0.70492      &0.16951        &9.4177     &8.4901     &9.8866     &22.12      &31.17\\
        2   &    &  &   &ShuffleNetV2
        &8.8042	    &9.09568	&0.651      &0.16491        &8.9307	    &8.2057	&9.2872 &11.815 &25.627\\
        3    &    &  &  &BlazeNet64 (Ours)
        &9.0876	&9.38528	&0.68196	&0.16739 &9.2175	&8.518	&9.6597  &11.000 &20.829\\
        \midrule
        4   &\multirow{8}*{AV-SkiM}       &\multirow{3}*{\cmark} &\multirow{3}*{25} &ResNet18  
        &8.9559	&9.2945	&0.7159	&0.1627 &8.7405	&8.9106	&9.4405 &19.146 &18.24\\
        5    &    &  & &ShuffleNetV2 
        &9.0467	&9.3479	&0.6949	&0.16591    &8.9507	&8.7535	&9.471    &8.8344   &12.693\\
        6    &    &  & &BlazeNet64 (Ours) 
        &9.1285	&9.42677	&0.7235	&0.1667    &8.922	&8.8683	&9.6597    &8.0565   &7.8969\\
        \cmidrule{3-14}
        7   &   &\multirow{2}*{\xmark}  &\multirow{2}*{25} &ResNet18  
        &12.8304	&13.0571	&1.07932	&0.208146    &13.1344	&12.1336	&13.3822   &32.88 &23.7\\ 
        8   &    &  &   &BlazeNet64 (Ours) 
        &12.8159	&13.0432	&1.0877	      &0.2078   &13.0608	&12.1393	&13.352     &21.79  &13.0\\
        \cmidrule{3-14}
        9   &   &\multirow{2}*{\cmark}  &12.5 &\multirow{2}*{BlazeNet64 (Ours)}  
        &6.3327	&6.6307	    &0.4649	&0.1209	    &6.2365	    &5.8156	    &6.7562	&8.0565   &6.6499\\ 
        10   &    &  &5   & 
        &0.0317	&0.0163	 &-0.2991	&0.00053	&-0.0089	&-0.0643	&0.0166	&8.0565   &5.9784\\
        \bottomrule
    \end{tabular}
    }
    \addtolength{\tabcolsep}{1.5pt}
    % \vspace*{-3mm}
    \label{tab:v_enc}
\end{table*}

\begin{table*}
    \centering
    \sisetup{
    detect-weight, % Make siunitx detect align bold cells correctly
    mode=text, % Make siuntix print tables in text mode (causes width of bold characters to be the same as non-bold)
    tight-spacing=true,
    round-mode=places,
    round-precision=2,
    table-format=1.2
    }
    \caption{The studies of our proposed acoustic encoder on the LRS3 datasets are detailed in the table. We present results for configurations with (\cmark) and without (\xmark) the acoustic encoder, as well as for systems trained with different loss functions. The SI-SNRi and SNRi values are reported in dB. All systems evaluated in this table utilize our proposed BlazeNet64 visual encoder. Bolded values highlight the best results obtained.}
    \vspace{-2mm}
    \addtolength{\tabcolsep}{-1pt}
    \resizebox{\linewidth}{!}{
    \begin{tabular}{cccc *{2}{S[round-precision=1,table-format=2.1]} SS *{5}{S[round-precision=1,table-format=2.1]}}
       \toprule
        \multirow{2}*{Sys}      &\multirow{2}*{Spk. Extractor} &\multirow{2}*{Acoustic Encoder}  &\multirow{2}*{Loss}  &{SI-SNRi}   &{SNRi}   &{PESQi}   &{STOIi} &{SI-SNRi} &{SI-SNRi} &{SI-SNRi}     &{\multirow{2}*{Params}}        &{\multirow{2}*{MACs}}\\
        &&&&{(all)}   &{(all)}   &{(all)}   &{(all)} &{(before)} &{(after)} &{(without)}\\
        \midrule
        6 &AV-SkiM  &\xmark  &$\mathcal{L}_{\text{SI-SNR}}$
        &9.1285	&9.42677	&0.7235	&0.1667    &8.922	&8.8683	&9.6597    &8.0565   &7.8969\\
        \cmidrule{1-13}  
        11 &\multirow{2}*{AV-SkiM-A} &\multirow{2}*{\cmark}  &$\mathcal{L}_{\text{SNR}}$
        &9.0794	&9.4017	&0.7205	&0.165   &9.0462	&8.8723	&9.3971 &8.5703    &8.923\\
        12  && &$\mathcal{L}_1 + \mathcal{L}_2$
        &\bfseries10.0475	&\bfseries10.35	&\bfseries0.82739	&\bfseries0.17923   &\bfseries9.8775	    &\bfseries9.8628	    &\bfseries10.5228   &8.5703 &8.923  \\
        \cmidrule{1-13}    
        13  &AV-SkiM  &\xmark &$\mathcal{L}_1$    
        &9.4537	&9.7435	&0.7462	&0.17506   &9.2812	&9.473	&9.7162  &8.0565  &7.8969\\
        \bottomrule
    \end{tabular}
    }
    \addtolength{\tabcolsep}{1pt}
    % \vspace*{-3mm}
    \label{tab:a_enc}
\end{table*}

\begin{table}
    \centering
    \sisetup{
    detect-weight, % Make siunitx detect align bold cells correctly
    mode=text, % Make siuntix print tables in text mode (causes width of bold characters to be the same as non-bold)
    tight-spacing=true,
    round-mode=places,
    round-precision=2,
    table-format=1.2
    }
    \caption{The SI-SNRi (dB) of models trained and evaluated on the VoxCeleb2 and TCD-TIMIT dataset mixtures.}
    \vspace{-2mm}
    % \addtolength{\tabcolsep}{-2.5pt}
    \resizebox{\linewidth}{!}{
    \begin{tabular}{cccc}
       \toprule
        Sys &Model                       &{VoxCeleb2}  & {TCD-TIMIT}\\
        \midrule
        14 &SkiM w/ ResNet18             &5.6    &9.1\\
        15 &SkiM w/ ShuffleNet           &5.5    &8.5 \\
        16 &SkiM w/ BlazeNet64 (Ours)                 &5.6    &9.5\\
        \midrule
        \multirow{2}*{17} &SkiM w/ BlazeNet64                  &\multirow{2}*{\bfseries6.4}   &\multirow{2}*{\bfseries11.0}\\
        &\& acoustic encoder (Ours)\\
        \bottomrule
    \end{tabular}
    }
    % \addtolength{\tabcolsep}{2.5pt}
    \vspace*{-3mm}
    \label{tab:more_results}
\end{table}

\subsection{Study on the acoustic encoder}

In table~\ref{tab:a_enc}, we evaluate the effectiveness of our proposed acoustic encoder. System 6 serves as the baseline, which does not include the acoustic encoder, whereas System 12 incorporates our proposed visual encoder along with the time-frequency-domain hybrid loss function. The results indicate that System 12 outperforms System 6 by $0.9$ dB in SI-SNRi, despite only a modest increase of $0.5$ million parameters and $1$ billion MACs. Notably, while the acoustic encoder is designed to leverage extraction momentum from past frames, it maintains performance stability even in scenarios where the target speaker changes. Specifically, the SI-SNRi values before and after a target change remain consistent at $9.9$ dB, demonstrating the robustness of the network in adapting to changes and effectively resetting the extraction momentum.

We employed the PARIS~\cite{pan2024autoregressive} training strategy for our autoregressive acoustic encoder. However, instead of the SNR loss function used in the original PARIS paper, we utilized the time-frequency-domain hybrid loss function ($\mathcal{L}_1 + \mathcal{L}_2$) for training. In our evaluation, System 11 represents a baseline trained with the SNR loss function alongside the acoustic encoder. It is observed that System 11 performs comparably to System 6, which is trained using SI-SNR, and does not surpass our proposed System 12. This result aligns with the general finding that systems trained with the SNR loss function typically do not achieve the same level of performance as those trained with the SI-SNR loss function. Conversely, our proposed System 12, which integrates SI-SNR loss with a frequency domain loss to regulate the output signal energy, demonstrates superior effectiveness.

We also present system 13 as an ablation study, which does not employ the acoustic encoder but is trained using the time-frequency-domain hybrid loss function $\mathcal{L}_1$, it has SI-SNRi of $9.5$ dB, which is better than system 6 but not better than system 12, showing the effectiveness of our acoustic encoder.

\subsection{Comparison on more datasets}
In Table~\ref{tab:more_results}, we compare our model and baselines on the VoxCeleb2~\cite{chung2018voxceleb2} and TCD-TIMIT dataset~\cite{harte2015tcd}, where the simulated mixtures are from ~\cite{pan2020muse}. On both dataset, our system 16 performs comparably or outperforms baseline systems 14 and 15 in terms of SI-SNRi. Our system 17 with an acoustic encoder performs the best with SI-SNRi of $6.4$ dB and $11.0$ dB on VoxCeleb2 and TCD-TIMIT mixtures, respectively.

\section{Conclusion}
In conclusion, this work presents a significant advancement in online audio-visual speaker extraction by addressing both computational efficiency and performance. The proposed visual encoder, with its lightweight design and efficient processing, provides a competitive alternative to the more complex visual encoder. The novel acoustic encoder, leveraging past frame information, effectively enhances the extraction process, and is robust to the dynamic scenarios involving target speaker changes. Overall, our approach demonstrates strong performance across various metrics with practical advantages in terms of computational efficiency for real-time applications in multi-talker environments.

\bibliographystyle{IEEEtran}
\bibliography{mybib}

\end{document}